\documentclass[prb,twocolumn,fleqn,showpacs]{revtex4}
\usepackage{graphicx}
\usepackage{amssymb}
\usepackage{amsmath}
\usepackage{bm}
\newcommand{\mb}   {\mathbf}      %
\newcommand{\mr}   {\mathrm}      %
\newcommand{\bra}  {\langle}      %
\newcommand{\ket}  {\rangle}      %
\newcommand{\up}   {\uparrow}     %
\newcommand{\dn}   {\downarrow}   %
\newcommand{\s}    {\sigma}       %
\newcommand{\w}    {\omega}       %

\newcommand{\eps}  {\epsilon} 
\begin{document}
\title{Quantum phase transitions in the honeycomb-lattice Hubbard model}
\author{K. Seki} \email{seki-kazuhiro@graduate.chiba-u.jp}
\author{Y. Ohta}
\affiliation{Department of Physics, Chiba University, Chiba 263-8522, Japan}
\date{\today}
\begin{abstract}
Quantum phase transitions in the Hubbard model on the honeycomb lattice 
are investigated in the variational cluster approximation.
The critical interaction for the paramagnetic to antiferromagnetic phase transition 
is found to be in remarkable agreement with a recent large-scale quantum Monte Carlo simulation.
Calculated staggered magnetization increases continuously with $U$ and 
thus we find the phase transition is of a second order.
We also find that the semimetal-insulator transition occurs at infinitesimally small interaction
and thus a paramagnetic insulating state appears in a wide interaction range.
A crossover behavior of electrons from itinerant to localized character
found  in the calculated single-particle excitation spectra and short-range spin correlation functions indicates that 
an effective spin model for the paramagnetic insulating phase is 
far from a simple Heisenberg model with a nearest-neighbor exchange interaction.
\end{abstract}
\maketitle

\section{INTRODUCTION}
The correlation induced metal-insulator transition in the half-filled Hubbard model 
is one of the fundamental topics in strongly correlated electron systems.\cite{Gebhard} 
The Hamiltonian of the Hubbard model is given as
\begin{equation}
H  =  -t \sum_{ij, \s} (c_{i\s}^\dag c_{j\s} + \mr{H. c.}) 
     + U \sum_{i}       n_{i\up} n_{i\dn} - \mu \sum_{i,\s} n_{i\s},
\end{equation}
where 
$t$ is the hopping integral between neighboring sites,
$U$ is the on-site Coulomb interaction strength, and 
$\mu$ is the chemical potential maintaining the average particle density at half filling.
This is achieved by setting  $\mu = U/2$ due to the presence of the particle-hole symmetry.
$c_{i\s}^\dag$ ($c_{i\s}$) denotes the creation (annihilation) operator 
of an electron with spin direction $\s$ ($= \up, \dn$) on the $i$-th site and 
$n_{i \s} = c_{i\s}^\dag c_{i\s}$ is the number operator. 
Hereafter we use $t = 1$ as the unit of energy.

On the square lattice, assuming the paramagnetic state, 
the critical Coulomb interaction for the metal-insulator transition ($U_c$)
is roughly equivalent to the band width, as  
has been estimated by various numerical methods such as the 
cluster-extended dynamical mean-field theory (CDMFT), \cite{Zhang,Park} 
variational cluster approximation (VCA), \cite{Balzer} and 
variational Monte Carlo method. \cite{Miyagawa} 
If the antiferromagnetism (AFM) is allowed, however, 
the system immediately goes into the antiferromagnetic insulator state 
by switching on $U$, i.e., $U_{\mr{AF}} = 0$,   
due to  the perfect nesting of the Fermi surface 
with the nesting vector $\mb{Q}=(\pi,\pi)$.

On the honeycomb lattice, situation is different from that of the square lattice.
The tight-binding band dispersion exhibits a semimetallic (SM) behavior 
with the Dirac cones at the corners of the hexagonal Brillouin zone (K and K' points) and 
vanishing density of states at the Fermi level. 
The perfect nesting of the Fermi surface is absent due to the point-like Fermi surface and 
thus the critical Coulomb interactions for the semimetal-insulator transition 
(SMIT) and AFM transition can differ from those of the square lattice. 

So far, the quantum phase transitions in the honeycomb lattice 
have been investigated by various theoretical and numerical studies.
A renormalization group analysis in the large-$N$ limit, \cite{Herbut} where $N$ is the number of fermion flavor,  
predicted that, if the analysis holds in the physical case $N=2$, 
the SM-AFM transition occurs at a finite $U/t$ value.
The Mott transition in a paramagnetic state has been studied by 
DMFT calculations, \cite{Tran, Jafari}
where the SMIT is shown to occur at relatively large interaction strength ($U_{c} \gtrsim 10$).
Since the honeycomb lattice is the two-dimensional lattice 
with the smallest coordination number (three neighbors),
spatial correlations neglected in DMFT should be taken into account and 
thus CDMFT studies have also been performed.
A finite-temperature CDMFT with continuous-time quantum Monte Carlo impurity solver 
anticipated $U_c=3.3$ at zero temperature limit. \cite{Wu}
Another finite-temperature CDMFT with exact-diagonalization solver by Liebsch~\cite{Liebsch} 
showed that the critical interaction is about $U_c=3-4$,
but the explicit value of $U_c$ was not found due to the difficulty from limitation of the energy resolution.
Recently, He and Lu~\cite{He} showed that 
the SMIT occurs at $U_c=0$ and AMF transition is of the first order
with the lower and upper critical Coulomb interactions 
$U_{\mr{AF}1} = 4.6$ and $U_{\mr{AF}2} = 4.85$.     
Quantum Monte Carlo (QMC) simulations have also been performed by various authors. \cite{Sorella, Furukawa, Meng, Sorella2} 
In particular, the discovery of a spin-liquid phase  
between $U_c=3.5$ and $U_{\mr{AF}} = 4.3$ by Meng \textit{et al.}, \cite{Meng} 
has attracted much attention and stimulated researches on this issue.
Recently, however, a large scale QMC up to 2592 sites 
by Sorella \textit{et al.} ruled out the existence of the spin liquid state 
and found a direct SM-AFM transition at $U_{\mr{AF}} = 3.76 \pm 0.04$. \cite{Sorella2}
All in all, critical behaviors near the SMIT and AFM of the Hubbard model on the honeycomb lattice 
are rather controversial than those on the square lattice.   

\begin{figure}[t]
\begin{center}
\includegraphics[width=10.5pc]{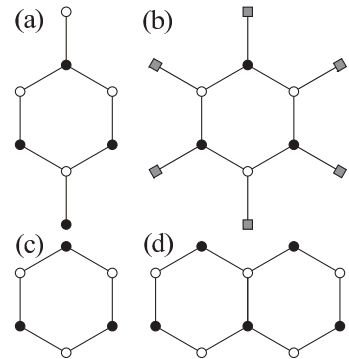}
\caption{
Clusters used in the present work.
(a) $L_c = 8$,
(b) $(L_c, L_b) = (6,6)$,
(c) $L_c =6$, and 
(d) $L_c=10$.
Filled (empty) circles represents the sites 
on the A (B)-sublattice of the honeycomb lattice and 
squares in (b) represents the bath sites. 
\label{fig0}}
\end{center}
\end{figure}

In this paper, motivated by such developments in the field, 
we study the AFM and SMIT in the Hubbard model on the honeycomb lattice 
by VCA with exact-diagonalization solver. \cite{Potthoff1, Potthoff2, Potthoffbook, SenechalReview}
We first calculate the AFM order parameter as a function of $U$ 
with various clusters such as $L_c= 6, 8, 10$ and $(L_c,L_b)=(6,6)$ clusters, where
$L_{c}$ and $L_{b}$ are the number of correlated sites and bath sites, respectively
(see Fig.\ref{fig0}). 
Then, to study the SMIT, 
the single-particle excitation spectra and single-particle gap are calculated as a function of $U$. 
We thus show that 
$U_{c}$ is in fact vanishingly small and 
therefore the paramagnetic insulating phase appears in a wide interaction strength.
We also calculate the spin correlation functions and 
find the development of localized moments, 
whereby the validity of effective spin models is discussed. 
We thus show that VCA is a suitable method to study the above issues
not only because it can take into account the spatial correlations important in low-dimensional systems, 
but also because it can be performed at zero temperature where the thermal fluctuations are absent and 
thus it can achieve a high energy-resolution essential for determining the SMIT.

\section{METHOD OF CALCULATION}
\subsection{Variational cluster approximation}
We employ the VCA\cite{Potthoff1, Potthoff2, Potthoffbook, SenechalReview} to study the SMIT and AFM 
of the Hubbard model on the honeycomb lattice at zero temperature.
In order to study AFM, we introduce the staggered magnetic field
\begin{equation}
H_{h'} = h' \sum_{i, \s} \mr{sign}(i) \s n_{i\s},
\end{equation}
where $\mr{sign}(i)= +1 (-1)$ for $i\in A(B)$-sublattice on the honeycomb lattice. 
Moreover, for the $(L_c,L_b) = (6,6)$ cluster, we introduce the bath hybridization
\begin{equation}
H_{V'} = V'        \sum_{i, \s} (c_{i\s}^\dag a_{i\s} + \mr{H. c.})
       + \eps_{a}' \sum_{i, \s}  a_{i\s}^\dag a_{i\s},
\end{equation}
where $a_{i\s}^\dag$ ($a_{i\s}$) denotes the creation (annihilation) operator of an electron with spin direction $\s$
on the $i$-th bath site and $V'$ is the hybridization between correlated sites and bath sites. 
The bath level is fixed at $\eps_{a}' = U/2$ 
so as to keep the average particle density of the correlated sites at half filling. 
Thus the fictitious magnetic field $h'$ and hybridization strength $V'$ are the variational parameters.
The eigenvalue problems for the Hamiltonian $H' = H + H_{h'} + H_{V'}$ defined on the small clusters are solved and 
the single-particle Green's functions are calculated by the Lanczos exact-diagonalization method.  
Then the grand-potential functional $\Omega(h',V')$ is calculated from the Green's functions and the stationarity 
point $(\partial \Omega/\partial h', \partial \Omega/ \partial V') = (0,0)$ is searched to find the physical solution.

Note that there are other possible variational parameters; e.g., 
enhancement of the hopping integrals between the correlated sites ($\Delta t$) and
the hopping integrals between the bath sites ($t_{\mr{bath}}$).
We have checked these variational parameters 
for some $U$ values on the $(L_c, L_b) = (6,6)$ cluster
and found that the optimal values are $|\Delta t| < 0.15$ and $|t_{\mr{bath}}| < 0.15$ and  
the optimal $V'$ values are affected only by $\sim 0.01|V'|$ 
in the presence of these $\Delta t$ or $t_{\mr{bath}}$, 
and moreover the change in the ground-state energy is negligible. 
Thus we omit the variations with respect to $\Delta t$ and $t_{\mr{bath}}$.
The similar discussion on the irrelevance of these variational parameters 
in the variational processes has been made 
by Balzer \textit{et al.}\cite{Balzer} 
for the half-filled Hubbard model on the square lattice.

\subsection{Cluster perturbation theory}
We use the cluster perturbation theory (CPT)\cite{Senechal1,Senechal2,Senechalbook} 
to obtain the 
lattice Green's function 
\begin{equation}
G_\mr{latt}^\alpha (\mb{k},\w) = 
\frac{1}{L_{\mr{c}}}  
\sum_{i,j\in \alpha}^{L_{\mr{c}}} 
G^{\mr{CPT}}_{ij}(\mb{k},\w) e^{-i \mb{k} \cdot (\mb{r}_i - \mb{r}_j)}.
\end{equation}
Here, 
$\alpha$ ($=A,B$) denotes the sublattice index,  
$\mb{r}_{i}$ is the position of the $i$-th site in the cluster, and 
\begin{equation}
{G}^{\mr{CPT}}(\mb{k},\w) = {G}'(\w) [I-{V}(\mb{k}){G}'(\w)] ^{-1}
\end{equation}
is the CPT Green's function,\cite{Senechal1, Senechal2,Senechalbook} where 
${G}'(\w)$ is the Green's function matrix of the cluster
and ${V}(\mb{k})$ represents the hopping matrix between the clusters.

\section{RESULTS OF CALCULATION}
\subsection{Staggered magnetization}
\begin{figure}[h]
\begin{center}
\includegraphics[width=16pc]{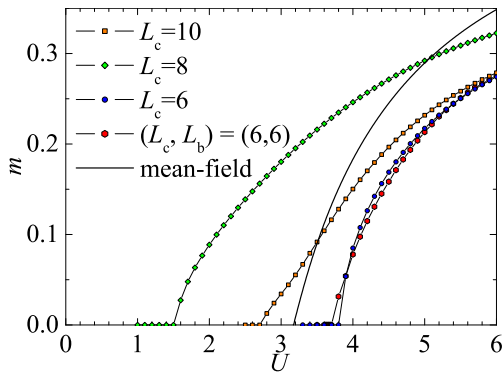}
\caption{
(Color online)
$U$ dependence of the staggered magnetization $m$ 
calculated for various clusters (see also Fig.~\ref{fig0}).
The mean-field result (solid line) is reproduced from Ref.~\onlinecite{Bercx}.
\label{fig1}}
\end{center}
\end{figure}

Figure~\ref{fig1} shows the $U$ dependence of the staggered magnetization defined as
\begin{equation}
m =  \frac{1}{2} \sum_{i\s} \mr{sign} (i) \s \bra n_{i\s} \ket \label{mag},
\end{equation}
where $\bra \cdots \ket$ denotes the ground-state expectation value
calculated by use of the CPT Green's function. \cite{SenechalReview}
A mean-field-theory (MFT) result reproduced from Ref.~\onlinecite{Bercx} is also shown.
We fine that $U_{\mr{AF}}$ strongly depends on the shape of the clusters, 
i.e., $U_{\mr{AF}} = $ 3.8, 1.5, 2.7, and 3.7
for $L_{c} = $ 6, 8, 10, and $(L_c, L_b) = (6,6)$, respectively. 
The order parameter increases continuously with increasing $U$ from $U_{\mr{AF}}$, 
and thus the transition is of the second-order.  
The characteristic linear increase of the order parameter $m \propto U-U_{\mr{AF}}$ near $U \approx U_{\mr{AF}}$
derived by the MFT~\cite{Sorella} is also observed in the VCA results.
Focusing on the results for the $L_c=6$ and $(L_c, L_b)=(6,6)$ clusters, 
we find that the bath degrees of freedom do not much affect the values of both $U_{\mr{AF}}$ and $m$.  
Our result $U_{\mr{AF}} = 3.7$ obtained for  the hexagonal cluster 
shows a remarkable agreement with the large-scale QMC simulation ($ U_{\mr{AF}} = 3.76 \pm 0.04$) \cite{Sorella2}
and reasonable agreement with the results of other previous studies. \cite{Sorella, Furukawa, Meng, He}
On the other hand, $U_{\mr{AF}}$ calculated for the $L_{c}=$ 8 and 10 clusters 
are lower than that obtained by the MFT ($U_{\mr{AF}} = 3.2$). 
Actually, being different from the $L_c=8$ and $10$ clusters, 
the hexagonal clusters do not suffer from boundary effects and 
spatial homogeneity is kept as it should be in the thermodynamic limit. 
Thus we conclude that $L_{c}=6$ and $(L_{c}, L_{b})=(6,6)$ clusters are  
the relevant choice of the clusters and hereafter we show the results of this choice. 

\subsection{Single-particle excitation}
\begin{figure}[t]
\begin{center}
\includegraphics[width=20pc]{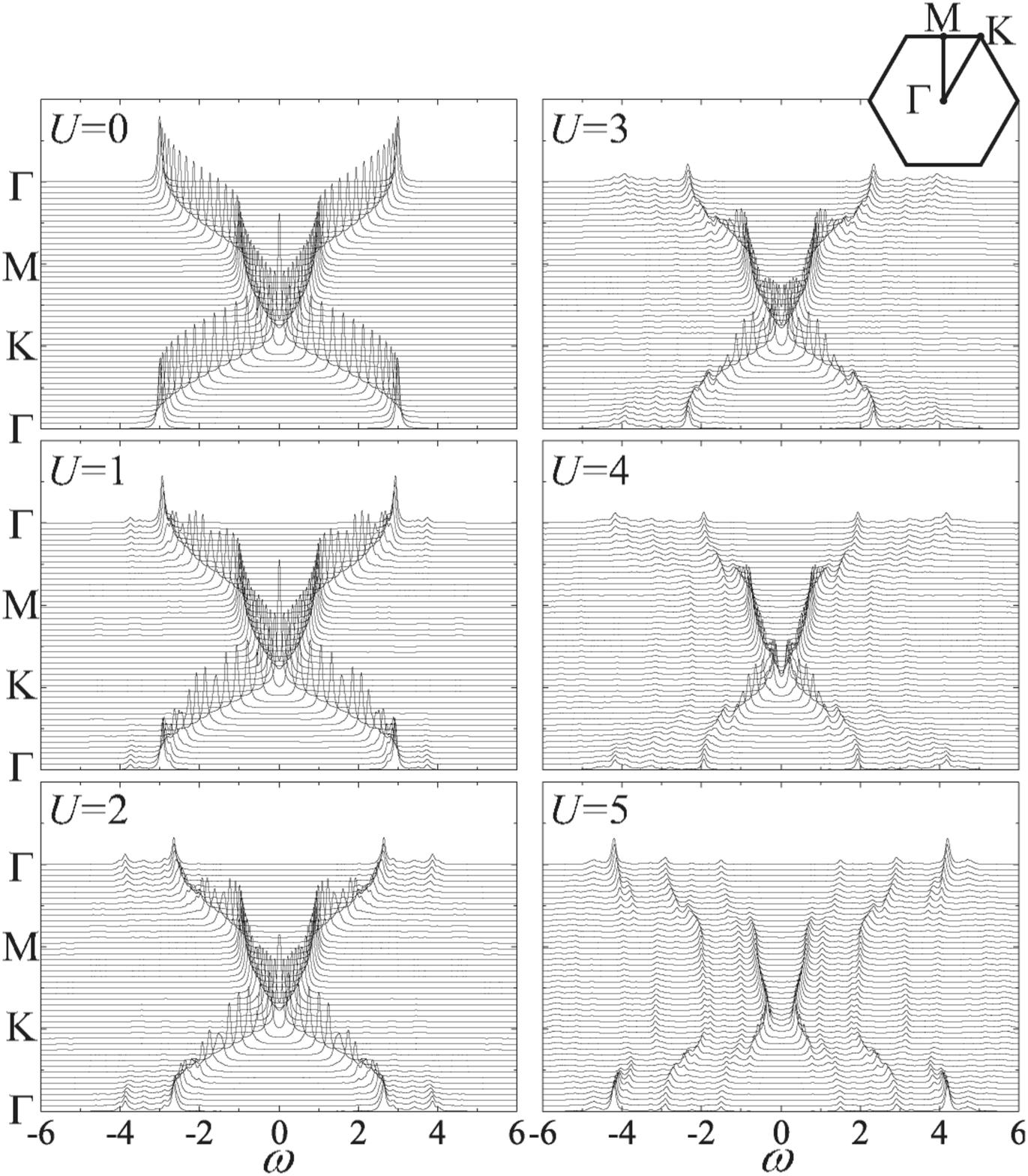}
\caption{
Single-particle excitation spectra $A(\mb{k},\w)$ calculated for a paramagnetic state.
Inset: definition of the $\Gamma$, M, and K points in the first Brillouin zone.
\label{fig2}}
\end{center}
\end{figure}

Next we calculate the single-particle excitation spectra defined as 
\begin{equation}
A(\mb{k},\w) = - \frac {1}{\pi}  \Im \sum_{\alpha = A,B} G_{\mr{latt}}^\alpha (\mb{k}, \w+i \eta), \label{Akw}
\end{equation}
where a small imaginary part $\eta$ of the complex frequency represents 
the Lorentzian broadening for the spectra.
We use $\eta = 0.1$.
The calculated results from $U=0$ to $5$ for a paramagnetic state 
with the $(L_c, L_b)=(6,6)$ cluster are shown in Fig.~\ref{fig2}.
At $U=4$, the spectra on the  $\Gamma$-K and $\Gamma$-M lines
become less dispersive, indicating the localized nature of electrons.  
On the contrary, on the K-M line (at the edge of the hexagonal Brillouin zone), 
the spectra remain to have quasiparticle-like sharp peaks even for $U \geq 4$.
At $U=5$, the upper and lower Hubbard bands appear clearly 
with a large intensity at $\w \sim \pm U$.
Although the single-particle gap exists clearly at $U = 5$,  
it is hard to judge when the gap opens by increasing $U$ 
due to the artificial Lorentzian broadening $\eta$. 
Thus we estimate the single-particle gap 
in a different way discussed in the next subsection.

\subsection{Single-particle gap}
The single-particle gap is defined as
\begin{equation}
\Delta = \mu^+  - \mu^-, \label{gap1}
\end{equation}
where $\mu^{+(-)}$ is the upper (lower) bound of the chemical potential.
Since $\mu^{+(-)}$ can be evaluated by calculating the particle density 
by integrating the CPT Green's function 
along the imaginary frequency axis, \cite{SenechalReview}
$\Delta$ does not suffer from the artificial Lorentzian broadening.

Another way to estimate the single-particle gap is 
to evaluate the intensity of the self-energy at the Fermi level. 
The lattice self-energy $\Sigma_{\mr{latt}} (\mb{k},\w)$ 
can be calculated from  $G^{-1}_{\mr{latt}}(\mb{k},\w)$ through the Dyson equation.
The intensity of the self-energy at the K point on the  Fermi level 
\begin{equation}
\sigma_{\mr{K}} =  -\w \Im \Sigma_{\mr{latt}} (\mr{K},i \w) |_{\w \rightarrow 0} 
\end{equation}
is related to the single-particle gap in the following way. 
It is known that the self-energy has only the simple poles 
on the real frequency axis. 
Thus, if the single-particle gap exists, the self-energy near the Fermi-level has the form
$\Sigma_{\mr{latt}} (\mr{K},\w) \approx \frac{\sigma_{\mr{K}}}{\w}$.
Therefore the lattice Green's function at the K point near the Fermi level is approximately given as
$
G_{\mr{latt}}(\mr{K},\w) \approx
\frac{1}{2} \left(
\frac{1}{\w - \sqrt{\sigma_{\mr{K}}} } + 
\frac{1}{\w + \sqrt{\sigma_{\mr{K}}} } \right),
$
i.e., the self-energy splits the spectrum above and below the Fermi level with the single-particle gap 
\begin{equation}
\Delta \simeq 2 \sqrt{\sigma_{\mr{K}}}. \label{gap2}
\end{equation}

\begin{figure}[t]
\begin{center}
\includegraphics[width=16pc]{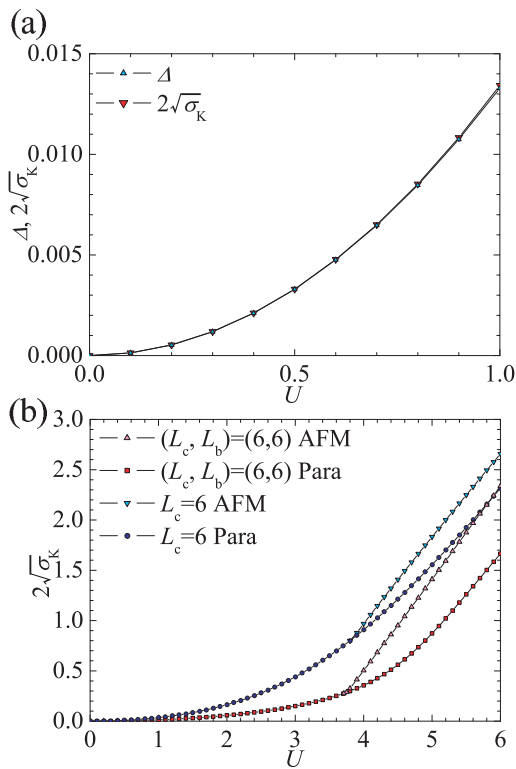}
\caption{
(Color online)
(a) 
$U$ dependence of the single-particle gap $\Delta$ 
calculated from Eq.~(\ref{gap1}) (triangles) and Eq.~(\ref{gap2}) (inverted triangles).
$(L_c,L_b)=(6,6)$ cluster is used for the calculation.
(b)
$U$ dependence of the single-particle gap from Eq.~(\ref{gap2}) 
calculated for $L_c=6$ and $(L_c,L_b)=(6,6)$ clusters.
\label{gapsmallU}}
\end{center}
\end{figure}

Figure \ref{gapsmallU} (a) shows 
the $U$-dependence of the single-particle gap estimated from 
the chemical potential difference Eq.~(\ref{gap1}) and 
intensity of the self-energy Eq.~(\ref{gap2}).
Although the gap is extremely small ($<0.001t$ for $U<0.2$), the two results nicely coincide.
We thus find that, even at $U=0.1$, a small but finite single-particle gap exists.
We may therefore conclude that the single-particle gap opens at infinitesimally small values of $U$,
just like in the one-dimensional Hubbard model at half filling, i.e., $U_c = 0$.
To see the effect of bath sites, 
we show the results for the $L_c=6$ and $(L_c, L_b) = (6,6)$ clusters in Fig.~\ref{gapsmallU} (b).
For $U \geq U_{\mr{AF}}$, results for both the AFM and paramagnetic states are shown.  
We find that the introduction of the bath sites significantly reduces the magnitude of the gap 
but it cannot close the gap, as we have seen in Fig. \ref{gapsmallU} (a).
For large $U$ regime, the gap increases linearly with $U$,
as it should be in the Mott insulator.

\subsection{Spin correlation function}
\begin{figure}[t]
\begin{center}
\includegraphics[width=16pc]{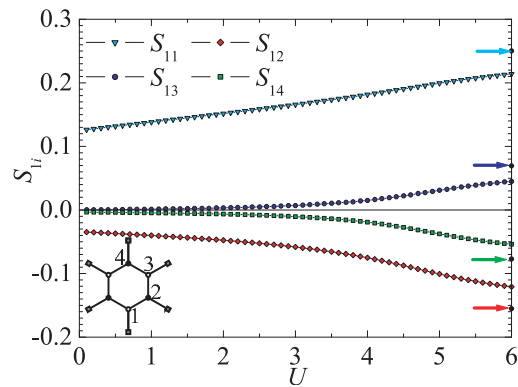}
\caption{
(Color online)
$U$ dependence of the spin correlation function of the $(L_c, L_b) = (6,6)$ cluster. 
Arrows indicate $S_{11}, S_{13}, S_{14}, $ and $S_{12}$ (from top to bottom) 
for a 6-site Heisenberg model with the nearest-neighbor exchange interaction.
Inset: definition of the site index $i$.}
\label{SiSj}
\end{center}
\end{figure}

Figure \ref{SiSj} shows the $U$ dependence of the spin correlation function in the cluster 
\begin{equation}
S_{1i}  =  \bra \Psi_{0} | S_{1}^z S_{i}^z |\Psi_{0} \ket,
\end{equation}
where $|\Psi_{0} \ket$ is the ground state of the $(L_c,L_b) = (6,6) $ cluster 
with the optimal hybridization $V'$
and $S_{i}^z = \sum_{\s} \s n_{i\s} /2$ is the spin operator for the correlated site $i$.
Positions of $i$-th sites are defined in the inset of Fig. \ref{SiSj}.
Calculations are done in the paramagnetic state.
For comparison, the same quantities for a 6-site Heisenberg model 
with the nearest-neighbor exchange interaction are shown by arrows. 

The on-site correlation function $S_{11}$ represents 
the development of the local moment. 
It increases monotonically and almost linearly with increasing $U$ 
from   $S_{11}=0.125$ in the non-interacting limit 
toward $S_{11}=0.25 $ in the  localized spin limit.
The second- and third-neighbor correlation functions 
$S_{13}$ and $|S_{14}|$ take small values for $U \lesssim 4$. 
This indicates that, although the single-particle gap is finite, 
electrons still keep their itinerancy and spin correlations
beyond the neighboring sites are not developed, 
unlike in the Heisenberg model.
Thus, to investigate the spin-liquid nature from an effective spin model,
the Heisenberg model with high $t/U$-order exchange interactions should  be necessary. 
Antiferromagnetic correlations between the neighboring sites $|S_{12}|$,
as well as $S_{13}$ and $|S_{14}|$, start increasing from $U \sim 4$, 
where the electrons begin to localize and local moments are developed.
This is consistent with the emergence of the AFM at $U_{\mr{AF}} = 3.7$ and 
the spectral signature of electron localization (see Fig.~\ref{fig2}). 
In this interaction regime, the Heisenberg model with the nearest-neighbor exchange interaction
may be appropriate for an effective low-energy model of the Hubbard model on the honeycomb lattice.

\section{SUMMARY}
We have studied the semimetal-insulator transition (SMIT) and 
antiferromagnetism (AFM)  in the half-filled Hubbard model on the honeycomb lattice 
within the variational cluster approximation (VCA).
The AFM transition was found to be of the second order and  
the critical Coulomb interaction for the AFM ($U_{\mr{AF}}$) 
obtained in the hexagonal clusters showed a remarkable agreement with the recent 
large-scale quantum Monte Carlo simulation. \cite{Sorella2}    
The single-particle gap has been calculated down to $U=0.1$ and found to be finite. 
Thus we concluded that the critical Coulomb interaction for the SMIT is $U_c=0$. 
 
The extremely small gap obtained by our zero-temperature calculations
in the small $U$ regime suggests that 
the careful evaluation of the single-particle gap is required 
for the SMIT on the honeycomb lattice. 
In the paramagnetic insulating state, 
we have calculated the spin-correlation functions 
and found that the local moments are not developed. 
Thus we concluded that, the high $t/U$-order exchange interactions should be necessary 
to investigate the spin-liquid state from effective spin models.
\section{ACKNOWLEDGMENTS}
We thank 
R. Eder and Y. Fuji 
for stimulating discussions.
K.S. acknowledges support from the JSPS Research Fellowship for Young Scientists.
This work was supported by Kakenhi Grant No. 22540363 of Japan.
A part of the computations was done at the research Center for Computational Science, Okazaki, Japan.


\begin{references}
\bibitem{Gebhard}  F. Gebhard, {\it The Mott Metal-Insulator Transition}, Springer Tracts in Modern Physics  Vol. 137 (Springer, Berlin, 1997).
\bibitem{Zhang}    Y. Z. Zhang and M. Imada,                                         {Phys. Rev. B}            {\bf 76}, 045108 (2007).	
\bibitem{Park}     H. Park, K. Haule, and G. Kotliar,                                {Phys. Rev. Lett.}        {\bf101}, 186403 (2008).
\bibitem{Balzer}   M. Balzer, B. Kyung, D. Senechal, A.-M. S. Tremblay, and M. Potthoff, {Europhys. Lett.}    {\bf 85}, 17002  (2009).
\bibitem{Miyagawa} T. Miyagawa and H. Yokoyama,                                   {J. Phys. Soc. Jpn.}      {\bf 80}, 084705 (2011).
\bibitem{Herbut}  I. F. Herbut,                                                       {Phys. Rev. Lett.}    {\bf 97}, 146401 (2006).
\bibitem{Tran}    M. T. Tran and K. Kuroki,                                          {Phys. Rev. B}    {\bf 79}, 125125  (2009). 
\bibitem{Jafari}  S. A. Jafari,                                                      {Eur. Phys. J. B} {\bf 68}, 537  (2009).   
\bibitem{Wu}      W. Wu, Y. H. Chen, H. S. Tao, N. H. Tong, and W. M. Liu,           {Phys. Rev. B}    {\bf  82}, 245102 (2010).
\bibitem{Liebsch} A. Liebsch,                                                     {Phys. Rev. B}    {\bf 83}, 035113  (2011).
\bibitem{He}      R. Q. He and Z. Y. Lu,                                          {Phys. Rev. B}    {\bf 86}, 045105  (2012).
\bibitem{Sorella}  S. Sorella and E. Tosatti                                       {Europhys. Lett.}     {\bf 19}, 699    (1992).
\bibitem{Furukawa} N. Furukawa,                                                   {J. Phys. Soc. Jpn.}  {\bf 70}, 1483   (2001).
\bibitem{Meng} Z.  Y. Meng, T. C. Lang, S. Wessel, F. F. Assaad, and A. Muramatsu, {Nature}             {\bf 464}, 08942  (2010). 
\bibitem{Sorella2} S. Sorella, Y. Otsuka, and S. Yunoki,                           e-print  arXiv:1207.1783v1.                             
\bibitem{Potthoff1}    M. Potthoff, M. Aichhorn, and C. Dahnken,            { Phys. Rev. Lett.}   {\bf 91},               206402 (2003).
\bibitem{Potthoff2}    M. Potthoff,                                         { Eur. Phys. J. B}    {\bf 32}, 429; {\bf 36},   335 (2003).
\bibitem{Potthoffbook} M. Potthoff, in {\it Strongly Correlated Systems: Theoretical Methods}, 
edited by A. Avella and F. Mancini, Springer Series in Solid-State Sciences, Vol. 171 (Springer, Berlin, 2012), pp. 303-339.
\bibitem{SenechalReview} D. S\'{e}n\'{e}chal,                              e-print arXiv:0806.2690v2.
\bibitem{Senechal1} D. S\'{e}n\'{e}chal, D. Perez, and M. Pioro-Ladriere, { Phys. Rev. Lett.}  {\bf 84},                  522 (2000).  
\bibitem{Senechal2} D. S\'{e}n\'{e}chal, D. Perez, and D. Plouffe,        { Phys. Rev. B}      {\bf 66},               075129 (2002).                   
\bibitem{Senechalbook} D. S\'{e}n\'{e}chal, in {\it Strongly Correlated Systems: Theoretical Methods}, 
edited by A. Avella and F. Mancini, Springer Series in Solid-State Sciences, Vol. 171 (Springer, Berlin, 2012), pp. 237-270.
\bibitem{Bercx} M. Bercx, T. C. Lang, and F. F. Assaad,                  { Phys. Rev. B}       {\bf 80},               045412 (2009).
\end{references}
\end{document}